\def \be{\begin{equation}}
\def \ee{\end{equation}}
\def \ba{\begin{eqnarray}}
\def \ea{\end{eqnarray}}

\newcommand{\CP}{{$\mathbb{C}${\rm P}}}
\newcommand{\Nm}{{${(N\!-\!1)}$\ }}

\documentclass[final,1p,times]{elsarticle}
\usepackage{color}
\usepackage{amssymb}
\journal{Elsevier}

\begin{document}

\begin{frontmatter}

\title{\CP\Nm Quantum Field Theories with Alkaline-Earth Atoms in Optical Lattices}

\author{C.~Laflamme$^{1,2}$, W.~Evans$^{3}$, M.~Dalmonte$^{1,2}$,  U.~Gerber$^{4,5}$, H.~Mej\'{i}a-D\'{i}az$^{4}$, W.~Bietenholz$^{4}$,  U.-J.~Wiese$^{3}$, P.~Zoller$^{1,2}$ }
\address{$^{1}$ Institute for Theoretical Physics, University of Innsbruck, A-6020, Innsbruck, Austria \\$^{2}$ Institute for Quantum Optics and Quantum Information, Austrian Academy of Sciences, A-6020, Innsbruck, Austria \\ $^{3}$ Albert Einstein Center for Fundamental Physics, Institute for Theoretical Physics, Bern University, 3012 Bern, Switzerland \\ $^{4}$ Instituto de Ciencias Nucleares, Universidad Nacional Aut\'{o}noma de M\'{e}xico, A.P.  70-543, C.P. 04510 Distrito Federal, Mexico \\
$^{5}$ Instituto de F\'{\i}sica y Matem\'{a}ticas, Universidad Michoacana de San Nicol\'{a}s de Hidalgo, Edificio C-3, Apdo.\ Postal 2-82,
C.P. 58040, Morelia, Michoac\'{a}n, Mexico}

\begin{abstract}

We propose a cold atom implementation to attain the continuum limit of $(1\!+\!1)$-d \CP\Nm quantum field theories. These theories share important features with $(3\!+\!1)$-d QCD, such as asymptotic freedom and $\theta$-vacua. Moreover, their continuum limit can be accessed via the mechanism of dimensional reduction. In our scheme, the \CP\Nm degrees of freedom emerge at low energies from a ladder system of SU($N$) quantum spins, where the $N$ spin states are embodied by the nuclear Zeeman states of alkaline-earth atoms, trapped in an optical lattice. Based on Monte Carlo results, we establish that the continuum limit can be demonstrated by an atomic quantum simulation by employing the feature of asymptotic freedom. We discuss a protocol for the adiabatic preparation of the ground state of the system, the real-time evolution of a false $\theta$-vacuum state after a quench, and we propose experiments to unravel the phase diagram at non-zero density.
\end{abstract}


\end{frontmatter}


\section{Introduction}
\label{}
Recently, there has been growing interest in developing physical platforms
for quantum simulation of Abelian and non-Abelian gauge theories~\cite{Banerjee12,Wiese:2013kk,Banerjee13,Zohar:2013eo,Tagliacozzo:2013bv,Stannigel:2014xy,Zohar:2015kx,Notarnicola:2015qy,Bazavov:2015zl}.
This effort is motivated by applications in particle and condensed
matter physics, with the hope of developing quantum simulation~\cite{Nascimbene12} as
a new tool to access regimes and phenomena complementary to, and beyond,
classical simulations~\cite{Wiese:2013kk}. Previous work has focused on implementing quantum
simulation of lattice gauge theories. An outstanding example
is provided by cold atoms in optical lattices as a natural and controlled
environment~\cite{Nascimbene12}, where the lattice gauge theory of interest emerges
as a low-energy effective description of tailored atomic Hubbard dynamics~\cite{Wiese:2013kk,Zohar:2015kx}. Applications
in particle physics, however, ultimately require taking the \emph{continuum
limit}, to eliminate artifacts due to space discretization. 


While some effective field theories emerge directly
from cold atom systems in continuous space (i.e. without a lattice), here we construct lattice
field theories from atoms in an optical lattice.
Instead of following the standard procedure of Wilson's lattice theory, where the continuum limit is approached by tuning a bare coupling
constant \cite{Wilson74}, we use the formalism of D-theory, in which the continuum
limit emerges via dimensional reduction \cite{Brower:1999kq,Brower2004149}.

We  illustrate this idea for the relevant example of \CP\Nm quantum
field theories \cite{ALV,Eichenherr1978215}. Such models have attracted interest
in the context of particle physics as toy models for QCD, with which
they share key features such as asymptotic freedom, the nonperturbative
generation of a mass gap, and the existence of nontrivial $\theta$-vacua \cite{ALV,Eichenherr1978215}. In addition, in a condensed matter context \CP\Nm models \cite{AuerbachBook} have been discussed in relation to deconfined quantum criticality \cite{Kaul:2012bs,Kaul:2015yg}. 

It has been shown that the (1+1)-d \CP\Nm model emerges via dimensional
reduction as the effective low-energy dynamics of certain (2+1)-d spin ladder
models of SU($N$) quantum magnetism \cite{Beard05}. In this paper we will show how this particular construction allows one to implement and approach the continuum limit of the \CP\Nm model in a natural and realistic way with fermionic Alkaline-Earth Atoms (AEAs) in an optical lattice~\cite{Takahashi07, Cazalilla:2009ff, Gorshkov10, Killian10,Schreck09,Schreck11,Takahashi11, Ye11, Takahashi12, Ye14, Fallani14,Foelling14,Cappellini:2014wu}.



\begin{figure}[t!]
\begin{center}
\includegraphics[width=0.9\columnwidth]{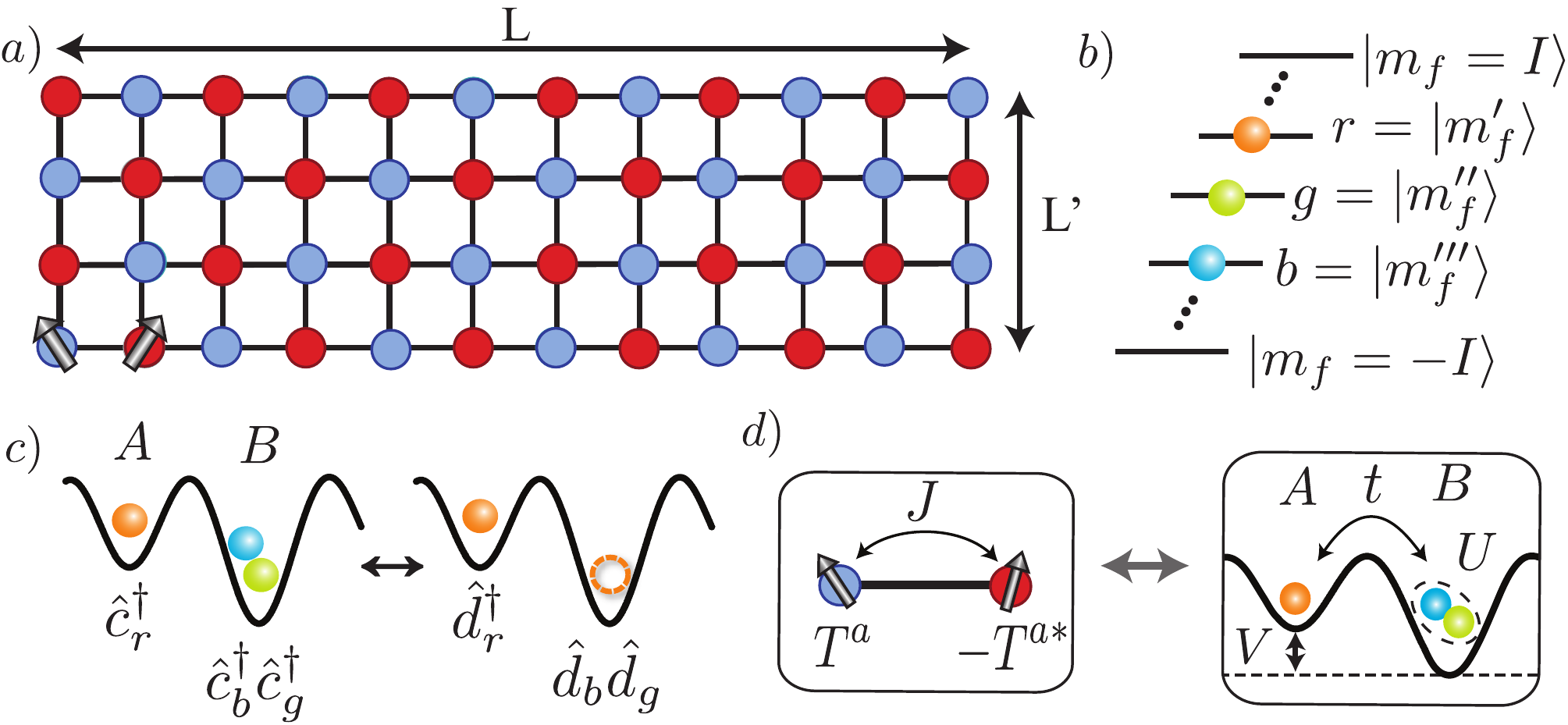}
\caption{a) Atomic setup for the implementation of the \CP\Nm model, where SU($N$) spins are arranged on a 2-d 
bipartite spatial lattice of volume $L \times L'$ with $L \gg L'$.  
b) The $N$ spin states are realized by AEAs occupying $N \leq 2I+1$ hyperfine states, where $I$ is the nuclear spin, with Zeeman splitting due to a uniform magnetic field. 
c) Identifying $c^\dagger \leftrightarrow d^\dagger$ (cf.\ Eqs.\ (\ref{eq:define_fundspins}) and (\ref{eq:Ham_Wannier})) on sublattice $A$, and $c^\dagger \leftrightarrow d$ on sublattice $B$, gives rise to the description of SU$(N)$ spins in the fundamental (red/dark), and anti-fundamental representation (blue/light), respectively, with 1 and $N-1$ fermions per site. d) Interactions between SU($N$) spins $T^a$ and $-T^{a*}$ are generated via superexchange, where $J\propto t^2/U$ .  An energy offset $V$ allows for full 
tunability of $J$.}
\label{fig:AF_image}
\vspace*{-5mm}
\end{center}
\end{figure}

The SU($N$) spin model of interest, and the corresponding atomic setup are illustrated in Fig.~\ref{fig:AF_image}. We assume fermionic AEAs with nuclear spin $I$ representing SU($N$) spins with $N\leq2I+1$. These atoms are loaded into a bipartite 2-d spatial lattice of volume $L \times L'$ ($L\gg L'$), realized as an optical superlattice. Such superlattices can be realized by superimposing a conventional optical lattice with lattice spacing $a$ with a second lattice of lattice spacing $\sqrt{2}a$ with a relative angle of $\pi/4$.The lattice depths and the interactions between the atoms are adjusted to achieve a filling with $1$ and $N-1$ atoms on the $A$ and $B$ sites of the bipartite lattice, respectively. Atoms on neighboring sites will then interact via superexchange (ie. Heisenberg-type) processes.  

Our paper is organized as follows. In Section \ref{sec:SUNmodel} we will begin with a description of the specific SU($N$) spin model of interest on a 2-d bipartite lattice. In Section \ref{sec:CPNmodel}, for self-consistency, we will review how the \CP\Nm model emerges as the low-energy effective theory of the specified SU($N$) spin model, following the discussion in Ref.~\cite{Beard05}. We will then discuss how this construction allows for the continuum limit of the (1+1)-d  \CP\Nm model to be approached via the mechanism known as dimensional reduction. In Section \ref{sec:implementation} we show how, by approaching the continuum limit in this way, we build a bridge which allows for their natural implementation via AEAs, while in Section~\ref{sec:imp} we present a summary on the possible experimental imperfections. Finally, in Sections \ref{sec:contlimit} to \ref{sec:QDynamics}, we describe the experimental signatures of the properties of the model, including asymptotic freedom and its relation to the continuum limit, the phase diagram of the system, and the quench dynamics of a false vacuum.

 \section{Anti-ferromagnetic SU($N$) Spin Model}

\label{sec:SUNmodel}

We begin by introducing the quantum spin model, from which the \CP\Nm model will be shown to emerge. As mentioned above, this model is defined on a 2-d bipartite lattice (see Fig.~\ref{fig:AF_image}), and described by the Hamiltonian
\be
\label{eq:AFchain_Ham}
H = J\sum_{\langle x y \rangle,\ x\in A} T^a_x \ \overline{T}^{a}_{y}, \qquad J>0,
\ee
where $T^{a}_{x}$ and $\overline{T}^{a}_{y}$ are generalized spin operators transforming under the fundamental and anti-fundamental representation of SU$(N)$, residing on sites $x \in A$ and $y \in B$ of the even and odd sublattices, respectively. The spin operators satisfy
$[T_{x}^{a},T_{y}^{b}] = {\rm i} \delta_{xy} f_{abc}T_{x}^{c}$, 
where $f_{abc}$ are the structure constants of SU($N$), and the anti-fundamental representation satisfies  $\overline{T}^{a}_{x}=-T^{a*}_{x}$. Note that in the case $N=2$ the fundamental and anti-fundamental representations are equivalent, and Eq.~(\ref{eq:AFchain_Ham}) reduces to the Heisenberg model. 

The system has a global SU($N$) symmetry with total spin conservation,
\be
[H,T^a]=\Big[ H,\sum_{x \in A} T_{x}^{a}-\sum_{y \in B} T_{y}^{a*} \Big] = 0 .
\ee 

The Hamiltonian can also be formulated in terms of fermionic operators, by rewriting the spins as fermionic bilinears
\be
\label{eq:define_fundspins}
T^{a}_{x} = \sum_{mm'}
 d^{\dagger}_{x m}  \lambda^{a}_{m m'} d_{x m'} , \ \
-T^{a*}_{x} = -\sum_{m m'} 
d^\dagger_{x m} \lambda^{a*}_{m m'} d_{x m'} ,
\ee 
where $\lambda^{a}$ are the generalized $N \times N$ Gell-Mann matrices,
${\rm Tr} [\lambda^{a} \lambda^{b} ] = 2 \, \delta^{ab}$, and $ d_{x m}$ annihilates a fermionic mode at position $x$, with the indices $m, m'\in\{1,\dots,N\}$ labelling $N$ fermionic states. Beyond being of direct interest for \CP\Nm models, this class of Hamiltonians has been extensively discussed in the context of frustrated magnetism as a natural extension of the conventional SU(2) Heisenberg model (see, e.g., Refs.~\cite{AuerbachBook,Kaul:2015yg}). In particular, in the 2-d limit, $L'\rightarrow \infty$, it has been shown how the SU($N$) symmetry undergoes spontaneous symmetry breaking~\cite{Sachdev89} for small $N$. We will exploit this property below while discussing the origin of the \CP\Nm quantum fields emerging from the spin Hamiltonian.

\section{Connection to the \CP($N-1$) Model}
\label{sec:CPNmodel}


In this section we will review how the (1+1)-d \CP\Nm model emerges as the effective low-energy dynamics of certain (2+1)-d SU($N$) spin models, showing how this theory undergoes dimensional reduction, approaching the continuum limit of the (1+1)-d CP($N$) model.  These results have been reported earlier \cite{Beard05}. However, to keep the paper self-contained we review them here, before discussing how this leads to a natural implementation with AEAs. 

The \CP\Nm models have been widely discussed in the context of low-dimensional quantum field theories. In contrast to their O($N$) counterparts, they display stable instanton solutions even for all $N$, as shown in Ref.~\cite{ALV}. This property, together with the fact that the models also show confinement and asymptotic freedom, are features  that (1+1)-d \CP\Nm models share with (3+1)-d QCD. 

While the \CP\Nm model can be studied analytically in the large $N$ limit \cite{ALV}, here we show the emergence of the model for $N=3,4$, allowing for quantum simulation complementary to analytical approaches. We begin with the SU($N$) spin Hamiltonian given in Eq.~(\ref{eq:AFchain_Ham}). 
In the zero-temperature thermodynamic limit $L, L' \rightarrow \infty$, the SU($N$) symmetry breaks spontaneously down to U($N-1$) \cite{Harada02}, 
resulting in $2(N-1)$ massless Nambu-Goldstone bosons described by fields in the coset space SU($N$)/U($N-1$) $\! =\,$\CP($N-1$). These fields can be described by $N\times N$ Hermitian projection matrices $P$, 
with $\mbox{Tr} \, P=1, P^{2}=P$ and $P^{\dagger} = P$, and by the action
\be
\label{eq:action}
S[P]= \int_0^\beta \! dt \int_0^L \! dx \int_0^{L'} \! dy \,  {\rm Tr}\big[\rho_s \partial_\mu P\partial_\mu P +\frac{\rho_s }{c^2}
\partial_tP\partial_t P - \frac{1}{2} (P\partial_x P \partial_t P -P\partial_t P \partial_x P  ) \big],
\ee
where $\rho_s$ is the spin stiffness parameter and $c$ is the spinwave velocity.
In the case of a finite extent $L'$, as a result of the Mermin-Wagner theorem, massless Nambu-Goldstone bosons are forbidden;
they pick up a mass $m=1/\xi$, where $\xi$ is the correlation length. Due to asymptotic freedom of the $(1+1)$-d \CP\Nm model, $\xi$ grows exponentially with $L'$,
\be
\label{eq:corr_length}
\xi \propto \exp (4 \pi L' \rho_{s} /(cN)).
\ee
As $L'$ increases, $\xi$ becomes much larger than $L'$, $\xi \gg L'$, and the fields become independent of the spatial direction $y$. In this regime, the system thus undergoes dimensional reduction, where the dynamics can then be described by the effective action in the remaining space-time dimension \cite{Beard05},
\be
\label{eq:action}
S[P]= \frac{c}{g^2}\!\!\int_0^\beta \! dt \,\int_0^L \! dx \,
{\rm Tr}\big[\partial_xP\partial_xP+\frac{1}{c^2}
\partial_tP\partial_t P\big] - {\rm i} \theta Q[P] , \
\ee
where $Q[P] \in \Pi_2[\mathbb{C}{\rm P}(N\!-\!1)] = \mathbb{Z}$ is the topological charge, and $g^2=c/(L'\rho_s)$ is the coupling constant of the dimensionally reduced theory. In terms of the lattice formulation, the vacuum angle is given by
$\theta=n\pi$ \cite{Beard05}, where $n=L'/a$  is the number of legs in the $L'$ direction, and $a$ is the lattice spacing. For $N=2$ this reduces to the well-known O(3) field theory description of the low-energy physics in Heisenberg antiferromagnets.

We can now highlight an important consequence of this construction. In contrast to Wilson's lattice field theory, in D-theory~\cite{WieseDtheory} the continuum limit, $\xi/a \rightarrow \infty$, is approached by increasing $L'$, not by decreasing a bare coupling constant. Due to the exponential dependence of $\xi$ on $L'=na$, cf.\ Eq.\ (\ref{eq:corr_length}), the continuum limit is already approached for moderate values of $L'$, which are accessible in current experiments. This strategy to regularize strongly coupled field theories is generally employed in the context of D-theory: in particular, in the D-theory regularization of QCD (3+1)-d gluon fields arise from dimensional reduction as collective excitations of (4+1)-d SU(3) quantum links, while chiral quarks arise as domain wall fermions~\cite{Brower:1999kq}.

\section{Implementation with AEAs}
\label{sec:implementation}

We now show how the construction of the \CP\Nm model presented in the previous section allows us to implement and observe properties of a quantum field theory approaching the continuum limit in an optical lattice. 

The Hamiltonian (\ref{eq:AFchain_Ham}) is realized in a natural way in a system of AEAs trapped in an optical lattice, based on their inherent SU$(N)$ symmetry \cite{Gorshkov10}.  Our implementation conceptually rests on two main ideas: first, using the formulation of the Hamiltonian (\ref{eq:AFchain_Ham}) in terms of fermionic degrees of freedom, as shown in Eq.~(\ref{eq:define_fundspins}), and second, implementing a particle-hole transformation to account for the fundamental/anti-fundamental representation of SU$(N)$ spins with $N\geq3$. In practice, this implementation exploits the toolbox already demonstrated in systems of trapped AEAs \cite{Takahashi07,Cazalilla:2009ff,Killian10,Schreck09,Gorshkov10,Schreck11,Takahashi11, Ye11, Takahashi12, Ye14, Fallani14,Foelling14}. A number of experiments with Sr and Yt atoms have already been realized, where various strongly correlated phases have been observed, including SU($N\leq6$) Luttinger liquids~\cite{Fallani14}, and SU($N$) Mott insulators~\cite{Takahashi12,Hofrichter:2015qy}.

To begin, we consider a system of fermionic AEAs trapped in a 2-d bipartite optical lattice, cf.\ Fig.\ \ref{fig:AF_image}.a). We assume the $2I+1$ nuclear spin states to be split in energy due to
 a uniform magnetic field (Fig.\ \ref{fig:AF_image}.b)). The Hamiltonian of such a system is expressed in terms of localized Wannier functions \cite{Gorshkov10} as
\ba
\label{eq:Ham_Wannier}
H &=& H_{t} + H_{U} ,   \nonumber \\ 
H_{t} &=& -t\sum_{m} \sum _{\langle x y\rangle}(c^{\dagger}_{x m} c_{y m} + c^{\dagger}_{y m} c_{x m}), \nonumber \\
H_{U} &=&  \frac{U}{2}\sum_{x}n_{x} (n_{x}-1) +V \sum _{x \in A} n_x.
\ea
Here $c_{x m}$ is the annihilation operator for an atom with nuclear spin $m\in\{-I,\dots,I\}$ in the Wannier function localized at site $x$, and $n_{x} = \sum_{m} c_{x m}^{\dagger} c_{x m}$ is the corresponding particle number operator. We denote by $t$ the 
nearest-neighbor hopping amplitude, $U$ is an on-site interaction energy, and $V$ is an energy offset between the two sublattices.
 Note that here the scattering length is independent of the nuclear spin level $m$, providing the system of AEAs with a global SU$(2I+1)$ symmetry. The reason is that the electronic and nuclear spin degrees of freedom are decoupled, implying that two atoms in different Zeeman states will intereact equally, independent of the value of $m_f$. This emergent SU($N$) symmetry has been shown to be accurate at a level of $10^{-8}$ \cite{Ye14}. We also note that one can take $N\leq 2I+1$ by initializing atoms into a subset of the magnetic states; if some Zeeman states are initially empty, their initial populations will not change due to the absence of spin-changing collisions. AEAs can realize SU$(N)$ physics up to $N=10$ (e.g. $^{87}$Sr and $^{173}$Yb), but here we will concentrate on the \CP(2) model, i.e.\ $N=3$, since it is particularly interesting from a theoretical viewpoint.\begin{figure}[t!]
\begin{center}
\includegraphics[width=0.9\textwidth]{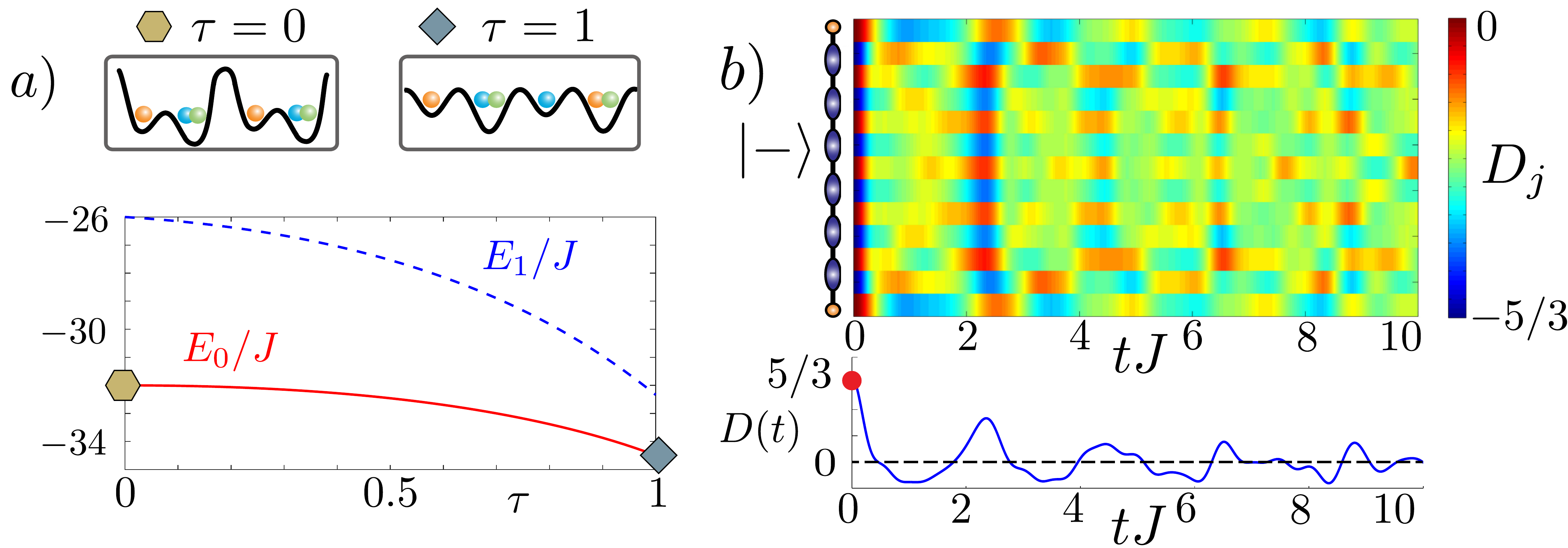}
\caption{Panel a) Energy of the ground state (red, solid) and excited state 
(blue, dashed), in units of the spin coupling $J$, as a function of the 
adiabatic parameter $\tau$. Schematics 
show the evolution of the system from $\tau=0$ (top left) to 
$\tau=1$ (top right). Panel b) Time evolution of the false vacuum state for a $14$ site system. Upper panel: Order parameter calculated at each bond as a function of time. Lower Panel: Time evolution of the order parameter summed over all bonds (see text).}
\label{fig:evolution}
\vspace*{-5mm}
\end{center}
\end{figure}
  
We initially occupy each site of the $A$ sublattice with $1$, and each site of the $B$ sublattice with $N-1$ atoms. While the $A$ sublattice spins in the fundamental representation of SU$(N)$ are embodied by a single fermion, the $B$ sublattice spins in the anti-fundamental representation are embodied by $N-1$ fermions, which are equivalent to a single hole, cf.\ Fig.\ \ref{fig:AF_image}.c). 
 Using the fermionic representation of Eq.\ (\ref{eq:define_fundspins}), we identify $c_x$ with $d_x$ on sublattice $A$, and with $d_x^\dagger$ on sublattice $B$.

We now consider the Hamiltonian (\ref{eq:Ham_Wannier}) in the strong coupling regime, $t \ll U, V$: the contribution of $H_{t}$ causes virtual tunneling processes within the subspace of states with fixed particle numbers per site (the eigenstates of $H_U$), thus generating SU$(N)$ superexchange terms \cite{Foelling14}. In this regime, to second order 
in $t/U$, the Hamiltonian (\ref{eq:AFchain_Ham}) emerges with \cite{AuerbachBook}
\be
\label{eq:Ham_eff}
J = \frac{t^{2} U}{(-V+U(N-3))(V-U(N-1))},
\ee
where an antiferromagnet requires $J>0$. 

The ground state of the system can be prepared via an adiabatic protocol: we start by preparing a band insulator with $N$ particles per site on a simple square lattice.  The population in each spin state can be controlled using, e.g.\ optical pumping \cite{Fallani14, Foelling14}. Each site is subsequently split into a double-well by adiabatically ramping up a superlattice, realizing a system of generalized SU$(N)$ singlets akin to what has already been realized using bosonic alkali atoms \cite{Bloch12}. 

The barrier between the wells is then adiabatically 
turned off, realizing the full quantum dynamics of Eq.\ (\ref{eq:AFchain_Ham}). This procedure relies entirely on existing techniques, and can be applied for $N=3,4$ and various $L'$. For a single chain this works as follows:
starting from a perfectly dimerized initial state, the inter-well exchange is switched on according to 
the time-dependent Hamiltonian
\be
H(\tau) = -(1-\tau)J\sum_{x\in A}T^a_xT^{a*}_{x+\hat{1}}+\tau H,
\ee
where $\tau \in [0,1]$ is the adiabatic parameter. The corresponding low-lying spectrum of a 14 site system is shown in Fig.\ \ref{fig:evolution} a). The system does not undergo a phase transition during this process; the gap does not close while changing $\tau$ from 0 to 1.~\footnote{We have numerically checked that the gap remains of order $0.6J$ even at $L=60$.} This ensures that an adiabatic ramp can be performed on time scales shorter than $1/J$.

\section{Experimental Imperfections}
\label{sec:imp}
In this subsection we analyse how robust our proposal is in light of the possible imperfections present in such systems of AEAs in an optical lattice. The main sources of imperfections are:
\begin{itemize}
\item effects of atom losses,
\item imperfect loading of the initial Mott phase,
\item role of external confinement and control over the number of legs,
\item spin population imbalance.
\end{itemize}
Below, we discuss in detail the relevance of each of these points for the observability of \CP\Nm physics within our scheme.

\subsection{Atom loss}

One of the most prominent sources of imperfection in an optical lattice experiment is the loss of atoms from the lattice. Such losses can occur both while preparing the initial Mott insulator phase, and throughout the evolution, due to, for example, three-body collisions. 

If we focus on $N=3$, sites occupied with 3 atoms only occur virtually, and thus three-body collisions are not a relevant source of imperfections. 

While sites with 3 atoms do occur at the initial stage of our adiabatic state preparation, the lattice at this point has twice the lattice spacing, which results in much more localized Wannier functions. \footnote{Assuming a loss rate of $K_3 \simeq 10^{-40}m^6/s$, typical for fermionic isotopes of AEAs \cite{0034-4885-77-12-124401}, an estimate of losses for a lattice spacing of $\sim 1 \mu m$ and a ramp time of $20$ms, is $0.2\%$ This is in contrast to $13\%$ for a lattice spacing at half of this ($\sim 500$nm).}

\subsection{Imperfect loading of the Mott state}

We consider in more detail defects in the initial Mott insulator phase, which arise in the preparation of the band insulator with $N$ atoms per site. Recent experiments have reported such errors on the level of $1$ missing atom per about $100$ sites for bosonic systems~\cite{Fukuhara:2015jk,Gross:2014qv}. Similar numbers have been reported for SU($2$) fermions \cite{foellingCommunication, Greif:2015aa}, while for SU($N$) ($N>2$), the additional effect of Pomeranchuk cooling would probably lead to an even smaller number of defects~\cite{Taie:2012ys} .

In our setup, this error implies the following: When we split each initialized site to create the required double-well structure, we do not have $1$ and $N-1$ atoms on each sublattice necessary to realize an SU($N$) spin in the fundamental and anti-fundamental representations, respectively. 

For concreteness we focus on $N=3$ in the remaining discussion. In this case, the Mott insulator phase should be initialized with $N=3$ atoms per site and the anti-ferromagnetic condition requires that $V<2U$. If one atom is missing, for fixed $U$, there are two possible configurations when the site is subsequently split into a double-well, corresponding to $V<U$ and $V>U$. In the first case, where $V<U$, one atom will be present on each site. In terms of the spin description of the system this imperfection corresponds to having a fundamental spin on a site where a spin in the anti-fundamental representation should be located. The spin-spin coupling between two spins of the same representation is $J = t^2U/(V^2-U^2)$. In the second case, where $V>U$, the ground state will have both atoms on the same site, with the other site empty. In the spin description, this corresponds to one spin missing from the system. 

In general, the system's dynamics in the presence of such imperfections would be described by a generalized SU($N$) $t$-$J$ model. However, for the regime $2U>V>U$ (which is experimentally reachable and satisfies the anti-ferromagnetic coupling constraint $J>0$ in Eq.\ (7)),  the dynamics of the system simplifies. In this case, the impurities are `static', as moving atoms in those partially filled sites is an off-resonant process. The dynamics of interest is then described by the SU($N$) spin ladder model, with a variable percentage of defects, depending on the initial number of missing atoms. We remark that the effects of such 'non-magnetic' impurities have been investigated in a related scenario for strongly frustrated magnets as well~\cite{Metlitski2008}.

\subsection{Role of external confinement and a `fuzzy' value of $L'$}
In our description of the system we have relied on two assumptions about its geometry, namely {\it i)} that the system is confined in a region of constant density, and {\it ii)} that $L'=na$ was given by a well-defined value of $n$, the number of legs in the $L'$ direction. The latter is crucial for a well-defined value of $\theta$ in the action of the \CP\Nm model, thus this point is essential for the proper implementation. 

In current experiments there are two ways to control 'sharp' boundaries. The first is to employ a box potential on the optical lattice which allows for a sharp edge of the system. For system sizes $\sim 50\mu$m the boundary effects will affect atoms within $\sim 2\mu$m of the box boundary~\cite{Chomaz:2015db,Navon:2015ul}. For Bose-Einstein condensates, the effects of such boundaries have been quantified to correspond to extremely flat potentials of the form $V(r) \propto ((r-r_0)/a)^\alpha$ with $\alpha>10$, and $r_0 \simeq 25\mu$m. The effect on the spin Hamiltonian is an adjusted coupling, $J$, between the spins, due to an effective chemical potential in the presence of the box potential --- an extremely small effect in our case. 

The second alternative is provided by the rapidly expanding technology of quantum gas microscopes (see, e.g., Ref.~\cite{Gross:2014qv} for a review). In this case, sharp boundaries in the system can be imprinted at the single-site level using blasting beams, so that the system dynamics is effectively confined into sharp boxes. The additional underlying confining potential will play no role in this case, as it does not affect the spin degrees of freedom.
 
\subsection{Spin population balance}
\label{sec:popbalance}
One further imperfection which could be present in the experimental realization is the initialization of the system with unequal numbers of atoms in each of the $N$ spin states. In reality, these spin states can be controlled via optical pumping, with current experiments achieving an accuracy of less than $1$ error in $\sim 100$ sites~\cite{Pagano:2014kq}. In the case of an error, the low-energy effective model is still the $\mathbb{C}$P($N-1$) model, however, with a non-zero particle number (see the discussion in Sec.~\ref{sec:FiniteDensity}). 
We remark that using spin-dependent in situ imaging (as in Ref.~\cite{Preiss:2015ve}) could also be useful as an efficient post-selection method to detect possible defects.

With this in mind, the possibility of selecting spin populations using optical pumping is actually available as an additional feature of our implementation: By intentionally loading an imbalance of spin states, we can investigate the $\mathbb{C}$P($N-1$) model at non-zero chemical potential. One can also employ Monte Carlo methods to investigate this problem, which will be the subject of future work.

\section{Continuum Limit}
\label{sec:contlimit}
\begin{figure}[t!]
\begin{center}
\includegraphics[width=0.9\textwidth]{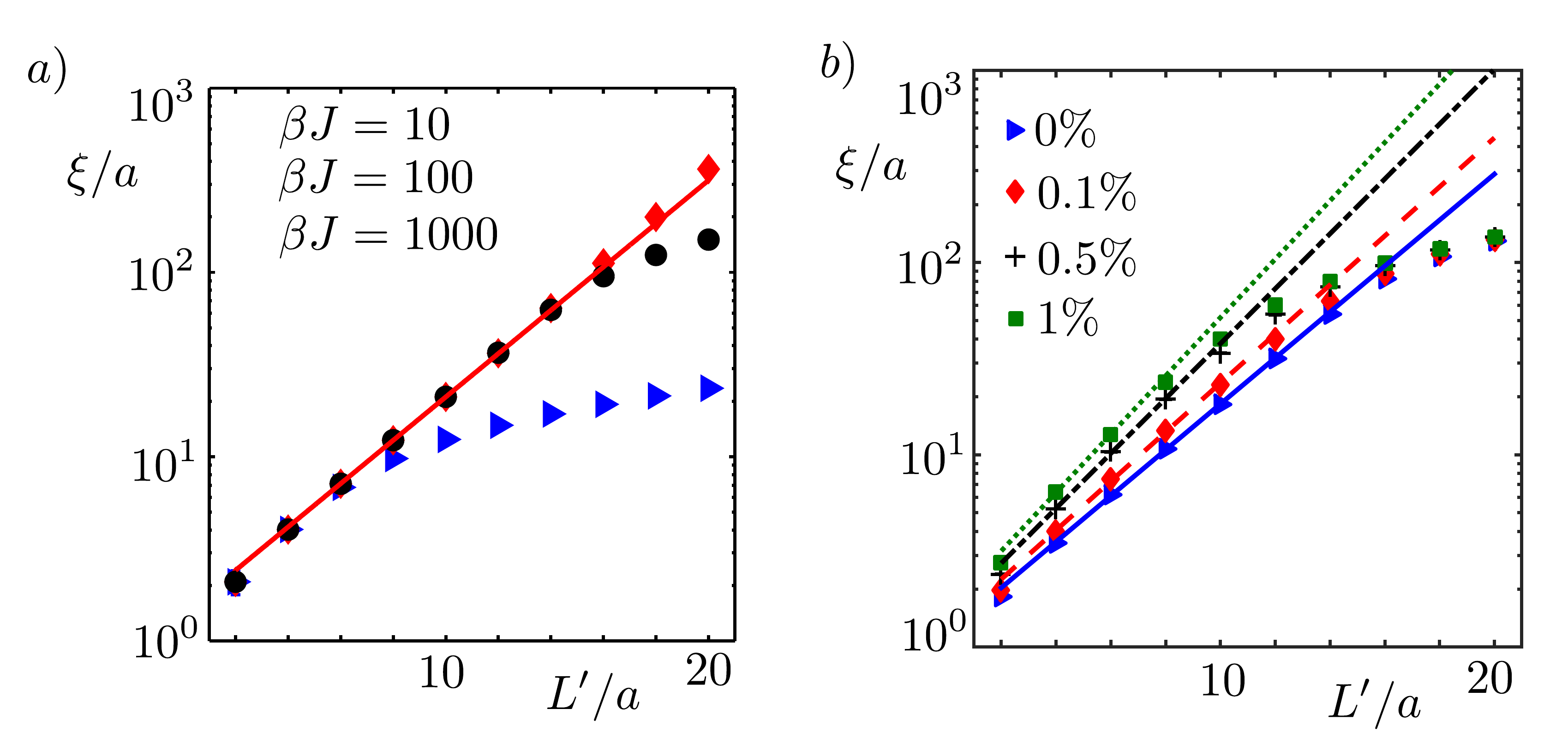}
\caption{Results for the correlation length $\xi$ (in the $L$-direction) from Monte Carlo simulations at  $L=1500a$. Panel a): Results at $\beta J=10$ (blue triangles), $\beta J=100$ (black circles) and $\beta J=1000$ (red diamonds). When the correlation length becomes comparable to the relevant energy scale set by the temperature it begins to deviate from the form in Eq.~(\ref{eq:corr_length}).  Panel b): Results for the correlation length $\xi$ at $\beta J=100$ in the presence of $0\%$ defects (blue triangles),  $0.1\%$ defects (red diamonds), $0.5\%$ defects (black crosses) and  $1\%$ defects (green squares).  
}
\label{fig:correlation}
\vspace*{-5mm}
\end{center}
\end{figure}

In order to demonstrate explicitly that the SU$(3)$ spin ladder gives rise to the \CP$(2)$ model in the continuum limit, it is vital to study the correlation length and verify that it increases exponentially with the size $L'$ of the extra dimension, cf. Eq.\ (\ref{eq:corr_length}). By means of Monte Carlo simulations with a loop cluster algorithm \cite{Haradaworldlines} we have calculated the spatial correlation length $\xi$. \footnote{The Monte Carlo simulations were performed with periodic boundary conditions in the $L$ direction, and open boundary conditions in the $L'$ direction. We measured the correlation length $\xi$ and the second moment
correlation length $\xi_2$ to ensure that their difference is
negligible.} 
For even $L'$ we obtain Fig.\ \ref{fig:correlation}.a), which indeed shows the anticipated exponential increase of $\xi$ with $L'$, in agreement with asymptotic freedom of the \CP$(2)$ model that emerges via dimensional reduction. The increase of the correlation length for $L'/a=4$ to $12$ should already be accessible using current experimental techniques. Even at temperatures around $\beta J \sim 10$, corresponding to a physical temperature of the order of nK, the correlation length $\xi(6a)$ is close to $10 a$, and already falls on the exponential that indicates asymptotic freedom.  The correlation length $\xi$ can be measured in a cold atoms setup via Bragg spectroscopy or through noise correlations \cite{Nascimbene12}.

\subsection{Continuum Limit in the presence of imperfections}
After the previous qualititative discussion on the effect of imperfections in the implementation with AEAs, here we consider quantitatively the effect of such imperfections on the observation of the correlation length $\xi$. 
The key question we would like to answer here is, whether or not asymptotic
freedom persists in the presence of such defects, and at which point the exponential growth of $\xi$ sets in. At a qualitative
level, one would expect that the phenomenon stays intact for low
concentrations, since a small number of defects will not drastically
affect the (2+1)-d limit, where spontaneous symmetry breaking occurs.
Nevertheless, numerical simulations have been performed in order to make
concrete statements regarding experimental realizations. We have performed a Monte Carlo study in which defects have been modelled as an empty lattice site
and 50 realizations of randomly distributed defects have been averaged over. We investigated concentrations of $0.1\%, 0.5\%$ and $1\%$ for $\beta J =
100$. In each case there is an $L'$ range in which the
exponential law demonstrating asymptotic freedom is clearly visible.

One can see from the results in Fig.~3.b) that the correlation length at a given $L'$ increases with the defect concentration, thus the regime where dimensional reduction occurs sets in earlier. As well, independent of the defect concentration there is a scale --- set by the inverse temperature --- at
which the correlation length saturates due to thermal fluctuations. This behavior is noticable in Fig.~3
both in the case with and without defects. Because the correlation length at a given $L'$ increases with the defect
concentration, this implies that the $L'$ range in which an exponential behaviour is visible
shrinks with the defect concentration. We have confirmed that the large $L'$
behaviour  of the curves in Fig 3.b) is a thermal effect by performing
some simulations at $\beta J =
1000$. 

Overall, the
only quantitative change defects cause is to renormalize the spin stiffness
of the system such that it increases with the defect concentration. This is
illustrated by the exponential fit lines in Fig 3.b), where one can see
the gradient of these lines increasing with defect concentration. The data are fitted up to values of $L'$ where this temperature effect becomes prominent which, for $\beta J =100$, is on the order of $\xi \sim 20a$.

\section{Finite Density Phase Diagram}
\label{sec:FiniteDensity}

Just like QCD, \CP\Nm models have a finite density phase diagram that is worth exploring. While in QCD a chemical potential $\mu$ can be coupled to the baryon number, in the \CP(2) model two chemical potentials, $\mu_3$ and $\mu_8$, can be coupled to the global SU$(3)$ symmetry, which is thereby explicitly broken down to U(1)$\times$U(1), or to SU(2)$\times$U(1) along the solid lines in Fig.\ \ref{fig:VBS_states}.a). It is then interesting to ask whether the U(1) symmetries are affected by Berezinskii-Kosterlitz-Thouless (BKT) transitions. In a cold atom experiment a finite density situation corresponds to loading the optical lattice with unequal numbers of atoms in the three Zeeman states, cf. Subsection~\ref{sec:popbalance}, corresponding to additional terms in the Hamiltonian,  
\be
H_{\mu}=-\mu_3 d^\dagger_x \lambda^3 d_{x} -\mu_8 d^\dagger_x \lambda^8 d_{x},
\ee
where $\lambda^{3}$ and $\lambda^{8}$ are the diagonal Gell-Mann matrices.
 A BKT transition is then signalled by the `condensation' of bosonic molecules, formed from two fermions (thus forming a bosonic pair) in specific combinations of Zeeman states, in the spirit of color superfluidity \cite{QCDbook,Rapp:2007ad,Kantian:2009qv}. 

\section{Spontaneous C-breaking at $\theta=\pi$}
\label{sec:Cbreaking}
\begin{figure}[t!]
\begin{center}
\includegraphics[width=0.9\columnwidth]{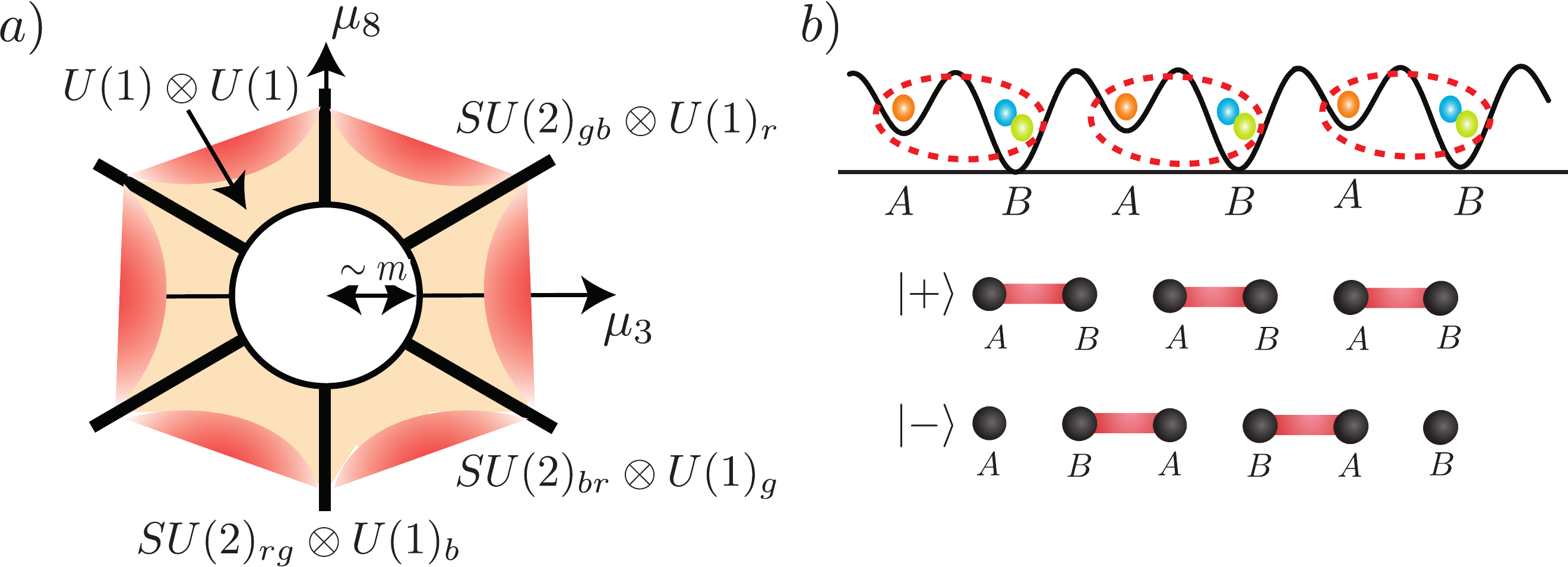}
\caption{Panel a) Conjectured phase diagram of the \CP$(2)$ model as a function of the chemical potentials $\mu_3,\mu_8$. In the vicinity of $\mu_3=\mu_8=0$, the system is in the vacuum state. Besides a normal fluid phase (yellow/light), high density color superfluid phases (red/dark) are expected to appear. The phase diagram can be explored in the proposed cold atom setup. Panel b) Symmetry of the ground states for $\theta = \pi$. Above:
In the ground state bonds emerge between two neighboring sites 
on sublattice $A$ and $B$. The ordering of the bonds gives rise to a 
double degeneracy of the ground state. Below:  Cartoon of 
the two degenerate ground states. A solid line indicates a dominant $\langle T^a_x T^{a*}_{x+\hat{1}}\rangle$, while no line represents a 
smaller value. }
\label{fig:VBS_states}
\vspace*{-5mm}
\end{center}
\end{figure}

Having advocated the feasibility of approaching the continuum limit in a quantum simulation, we now consider an odd number $n$ of transversely coupled chains, corresponding to $\theta = \pi$. 
At this point, analytical considerations suggest a first order phase 
transition with spontaneous charge conjugation (C) symmetry breaking \cite{Seiberg:1984eu}, which has been confirmed numerically \cite{ Beard05}. In our proposed experimental realization with discrete spins, C corresponds to a shift by one lattice spacing in the longitudinal $1$-direction, $T^a_x \rightarrow -T^{a*}_{x+\hat{1}}$,  $-T^{a*}_{x} \rightarrow T^{a}_{x+\hat{1}}$, 
where the sites $x$ and $x+\hat{1}$ belong to the
$A$ and $B$ sublattice, respectively. An order 
parameter which signals C-breaking in spin systems is 
given by \cite{Sachdev89} 
\be
\label{eq:order_param}
D =\sum_{x\in A}\langle T^a_xT^{a*}_{x+\hat{1}}- T^a_xT^{a*}_{x-\hat{1}}\rangle ,
\ee
which, equivalently, detects dimerization. 
When C is preserved ($n$ even, $\theta=0$) $D$ vanishes, whereas when it is spontaneously broken ($n$ odd, $\theta=\pi$) there are two degenerate ground states with opposite non-zero values of $D$, see Fig.\ \ref{fig:VBS_states}.
In a cold atom setup, measuring the singlets contributing to $D$ has been proposed and demonstrated via spin-changing collisions 
\cite{Paredes08, Nascimbene12}. A possible adiabatic preparation scheme for 
the realization of such generalized resonating valence bond states is illustrated in Fig.\ \ref{fig:evolution}.a).

\section{ Quench Dynamics Decay of a False Vacuum}
\label{sec:QDynamics}
The possibility of initializing states composed of singlets provides an opportunity to investigate real-time quenched dynamics driven by $H$. In a finite system with open boundary conditions and even $L$, one can investigate how a false vacuum $|-\rangle$, cf.\ Fig.\ \ref{fig:VBS_states}.b), decays as a function of time after the Hamiltonian $H$ is switched on. Such a false vacuum decay can mimic processes in inflationary early universe cosmology, as well as bubble nucleation at a first order phase transition. Due to energy conservation, the false vacuum, which has an energy cost for any finite system size $L$, cannot decay fully into the true vacuum $|+\rangle$. Instead one expects damped coherent oscillations between the two vacuum states. 
In order to quantify the false vacuum decay in real time, for a single chain we consider the order parameter \cite{Sachdev89} 
\be
 D(t) = \sum_{x\in A}\langle \Psi(t)\vert (T^a_xT^{a*}_{x+\hat{1}}- T^a_xT^{a*}_{x-\hat{1}})\vert \Psi(t)\rangle, 
 \ee
  which indicates whether singlet states predominantly form on the even or odd bonds. This order parameter is maximal for $|-\rangle$ (all even bonds have a singlet). In Fig.\ \ref{fig:evolution}.b), we show the dynamics of the false vacuum decay evaluated by exact diagonalization of a $L/a=14$ site system starting in the initial state $|-\rangle$. At times $t \ll1/J$ the even singlets (blue) are stable. At later times, the false vacuum decays, with correlations remaining only in the central part of the system, while the bonds close to the boundary revert the order. The decay of the full order parameter is depicted in the lower panel, which indeed shows coherent oscillations. Both the order parameter and the local singlet projectors can be experimentally measured as discussed in the previous section. While moderate system sizes can be reached using exact diagonalization, the real-time dynamics in the continuum limit is inaccessible to classical simulations. Experiments using the present scheme would shed light on the real-time dynamics of false vacua in \CP\Nm models.

\section{Conclusions}

We have outlined a proposal for the quantum simulation of a \CP\Nm quantum field theory using cold atoms trapped in an optical lattice with a ladder geometry. Our work shows how the continuum limit can be assessed using dimensional reduction, and how paradigmatic phenomena such as asymptotic freedom can be observed in cold atom experiments. Extending such investigations to non-Abelian gauge theories would provide an indispensable tool for the quantum simulation of fundamental theories such as QCD at finite baryon density.

\section{ Acknowledgments}
We thank D. Banerjee, L. Fallani, C.V. Kraus, M. Punk, E. Rico, and S. Sachdev for useful discussions. 
This work was supported by the Schweizerischer Nationalfonds, 
the European Research Council by means of the European Union's 
Seventh Framework Programme (FP7/2007-2013)/ERC grant agreement 339220,
the Mexican Consejo Nacional de Ciencia y Tecnolog\'{\i}a 
(CONACYT) through projects CB-2010/155905 and CB-2013/222812,
and by DGAPA-UNAM, grant IN107915.
Work in Innsbruck is partially supported by the ERC Synergy Grant UQUAM, SIQS, and the SFB FoQuS (FWF Project No. F4016-N23). 
CL was partially supported by NSERC.



  \bibliographystyle{elsarticle-num} 
    \bibliography{CPNRefs}

\begin{thebibliography}{10}
\expandafter\ifx\csname url\endcsname\relax
  \def\url#1{\texttt{#1}}\fi
\expandafter\ifx\csname urlprefix\endcsname\relax\def\urlprefix{URL }\fi
\expandafter\ifx\csname href\endcsname\relax
  \def\href#1#2{#2} \def\path#1{#1}\fi

\bibitem{Banerjee12}
D.~Banerjee, M.~Dalmonte, M.~M\"uller, E.~Rico, P.~Stebler, U.-J. Wiese,
  P.~Zoller, Phys. Rev. Lett. 109 (2012) 175302.

\bibitem{Wiese:2013kk}
U.-J. Wiese, Ann. Phys. 525~(10-11) (2013) 777--796.

\bibitem{Banerjee13}
D.~Banerjee, M.~B\"ogli, M.~Dalmonte, E.~Rico, P.~Stebler, U.-J. Wiese,
  P.~Zoller, Phys. Rev. Lett. 110 (2013) 125303.

\bibitem{Zohar:2013eo}
E.~Zohar, J.~I. Cirac, B.~Reznik, Phys. Rev. Lett. 110~(12) (2013) 125304.

\bibitem{Tagliacozzo:2013bv}
L.~Tagliacozzo, A.~Celi, A.~Zamora, M.~Lewenstein, Ann. Phys. 330 (2013)
  160--191.

\bibitem{Stannigel:2014xy}
K.~Stannigel, P.~Hauke, D.~Marcos, M.~Hafezi, S.~Diehl, M.~Dalmonte, P.~Zoller,
  Phys. Rev. Lett. 112 (2014) 120406.

\bibitem{Zohar:2015kx}
E.~Zohar, J.~I. Cirac, B.~Reznik, Rep. Prog. Phys. 79 (2016) 014401.

\bibitem{Notarnicola:2015qy}
S.~Notarnicola, E.~Ercolessi, P.~Facchi, G.~Marmo, S.~Pascazio, F.~V. Pepe, J.
  Phys. A 48 (2015) 30FT01.

\bibitem{Bazavov:2015zl}
A.~Bazavov, Y.~Meurice, S.-W. Tsai, J.~Unmuth-Yockey, J.~Zhang, Phys. Rev. D 92
  (2015) 076003.

\bibitem{Nascimbene12}
I.~Bloch, J.~Dalibard, S.~Nascimb{\`e}ne, Nat. Phys. 8~(4) (2012) 267--276.

\bibitem{Wilson74}
K.~G. Wilson, Phys. Rev. D 10 (1974) 2445--2459.

\bibitem{Brower:1999kq}
R.~Brower, S.~Chandrasekharan, U.-J. Wiese, Phys. Rev. D 60~(9) (1999) 094502.

\bibitem{Brower2004149}
R.~Brower, S.~Chandrasekharan, S.~Riederer, U.-J. Wiese, Nucl. Phys. B
  693~(1--3) (2004) 149 -- 175.

\bibitem{ALV}
A.~D'Adda, M.~L{\"u}scher, P.~Di~Vecchia, Nucl. Phys. B 146~(1) (1978) 63--76.

\bibitem{Eichenherr1978215}
H.~Eichenherr, Nucl. Phys. B 146~(1) (1978) 215 -- 223.

\bibitem{AuerbachBook}
A.~Auerbach, Interacting Electrons and Quantum Magnetism, Springer, 1994.

\bibitem{Kaul:2012bs}
R.~K. Kaul, A.~W. Sandvik, Phys. Rev. Lett. 108~(13) (2012) 137201.

\bibitem{Kaul:2015yg}
R.~K. Kaul, M.~Block, J. Phys.: Conf. Ser. 640 (2015) 012041.

\bibitem{Beard05}
B.~B. Beard, M.~Pepe, S.~Riederer, U.-J. Wiese, Phys. Rev. Lett. 94 (2005)
  010603.

\bibitem{Takahashi07}
T.~Fukuhara, Y.~Takasu, M.~Kumakura, Y.~Takahashi, Phys. Rev. Lett. 98 (2007)
  030401.

\bibitem{Cazalilla:2009ff}
M.~A. Cazalilla, A.~F. Ho, M.~Ueda, New Journal of Physics 11 (2009) 103033.

\bibitem{Gorshkov10}
A.~V. Gorshkov, M.~Hermele, V.~Gurarie, C.~Xu, P.~Julienne, J.~Ye, P.~Zoller,
  E.~Demler, M.~D. Lukin, A.~M. Rey, Nat. Phys. 6~(4) (2010) 289--295.

\bibitem{Killian10}
B.~J. DeSalvo, M.~Yan, P.~G. Mickelson, Y.~N. Martinez~de Escobar, T.~C.
  Killian, Phys. Rev. Lett. 105 (2010) 030402.

\bibitem{Schreck09}
S.~Stellmer, M.~K. Tey, B.~Huang, R.~Grimm, F.~Schreck, Phys. Rev. Lett. 103
  (2009) 200401.

\bibitem{Schreck11}
S.~Stellmer, R.~Grimm, F.~Schreck, Phys. Rev. A 84 (2011) 043611.

\bibitem{Takahashi11}
S.~Sugawa, K.~Inaba, S.~Taie, R.~Yamazaki, M.~Yamashita, Y.~Takahashi, Nat.
  Phys. 7~(8) (2011) 642--648.

\bibitem{Ye11}
M.~D. Swallows, M.~Bishof, Y.~Lin, S.~Blatt, M.~J. Martin, A.~M. Rey, J.~Ye,
  Science 331~(6020) (2011) 1043--1046.

\bibitem{Takahashi12}
S.~Taie, R.~Yamazaki, S.~Sugawa, Y.~Takahashi, Nat. Phys. 8~(11) (2012)
  825--830.

\bibitem{Ye14}
X.~Zhang, M.~Bishof, S.~L. Bromley, C.~V. Kraus, M.~S. Safronova, P.~Zoller,
  A.~M. Rey, J.~Ye, Science 345~(6203) (2014) 1467--1473.

\bibitem{Fallani14}
G.~Pagano, M.~Mancini, G.~Cappelline, P.~Lombardi, F.~Sch{\"a}fer, H.~Hu, X.-J.
  Lio, J.~Catani, C.~Sias, M.~Inguscio, L.~Fallani, Nat. Phys. 10~(3) (2014)
  198--201.

\bibitem{Foelling14}
F.~Scazza, C.~Hofrichter, M.~H{\"o}fer, P.~C. De~Groot, I.~Bloch,
  S.~F{\"o}lling, Nat. Phys. 10~(10) (2014) 779--784.

\bibitem{Cappellini:2014wu}
G.~Cappellini, M.~Mancini, G.~Pagano, P.~Lombardi, L.~Livi, M.~Siciliani~de
  Cumis, P.~Cancio, M.~Pizzocaro, D.~Calonico, F.~Levi, C.~Sias, J.~Catani,
  M.~Inguscio, L.~Fallani, Phys. Rev. Lett. 113~(12) (2014) 120402.

\bibitem{Sachdev89}
N.~Read, S.~Sachdev, Phys. Rev. Lett. 62 (1989) 1694--1697.

\bibitem{Harada02}
K.~Harada, N.~Kawashima, M.~Troyer, Phys. Rev. Lett. 90 (2003) 117203.

\bibitem{WieseDtheory}
U.-J. Wiese, PoS LAT2005 (2005) 281.

\bibitem{Hofrichter:2015qy}
C.~Hofrichter, L.~Riegger, F.~Scazza, M.~H{\"o}fer, D.~R. Fernandes, I.~Bloch,
  S.~F{\"o}lling, arXiv:1511.07287 [cond-mat.quant-gas].

\bibitem{Bloch12}
S.~Nascimb\`ene, Y.-A. Chen, M.~Atala, M.~Aidelsburger, S.~Trotzky, B.~Paredes,
  I.~Bloch, Phys. Rev. Lett. 108 (2012) 205301.

\bibitem{0034-4885-77-12-124401}
M.~A. Cazalilla, A.~M. Rey, Rep. Prog. Phys. 77~(12) (2014) 124401.

\bibitem{Fukuhara:2015jk}
T.~Fukuhara, S.~Hild, J.~Zeiher, P.~Schau{\ss}, I.~Bloch, M.~Endres, C.~Gross,
  Phys. Rev. Lett. 115 (2015) 035302.

\bibitem{Gross:2014qv}
C.~Gross, I.~Bloch, arXiv:1409.8501 [cond-mat.quant-gas].

\bibitem{foellingCommunication}
S.~F{\"o}lling, private communication.

\bibitem{Greif:2015aa}
D.~Greif, M.~F. Parsons, A.~Mazurenko, C.~S. Chiu, S.~Blatt, F.~Huber, G.~Ji,
  M.~Greiner, arXiv:1511.06366.

\bibitem{Taie:2012ys}
S.~Taie, R.~Yamazaki, S.~Sugawa, Y.~Takahashi, Nat. Phys. 8 (2012) 825--830.

\bibitem{Metlitski2008}
M.~A. Metlitski, S.~Sachdev, Phys. Rev. B 77~(5) (2008) 054411.

\bibitem{Chomaz:2015db}
L.~Chomaz, L.~Corman, T.~Bienaim{\'e}, R.~Desbuquois, C.~Weitenberg,
  S.~Nascimb{\`e}ne, J.~Beugnon, J.~Dalibard, Nat. Commun. 6 (2015) 6162.

\bibitem{Navon:2015ul}
N.~Navon, A.~L. Gaunt, R.~P. Smith, Z.~Hadzibabic, Science 347 (2015) 167--170.

\bibitem{Pagano:2014kq}
G.~Pagano, M.~Mancini, G.~Cappellini, P.~Lombardi, F.~Sch{\"a}fer, H.~Hu, X.-J.
  Liu, J.~Catani, C.~Sias, M.~Inguscio, L.~Fallani, Nat. Phys. 10 (2014)
  198--201.

\bibitem{Preiss:2015ve}
P.~M. Preiss, R.~Ma, M.~E. Tai, J.~Simon, M.~Greiner, Phys. Rev. A 91~(4)
  (2015) 041602(R).

\bibitem{Haradaworldlines}
N.~Kawashima, K.~Harada, JPSJ 73~(6) (2004) 1379--1414.

\bibitem{QCDbook}
K.~Rajagopal, F.~Wilczek, At the frontier of particle physics: Handbook of QCD,
  World Scientific, 2001.

\bibitem{Rapp:2007ad}
{\'A}.~Rapp, G.~Zarand, C.~Honerkamp, W.~Hofstetter, Phys. Rev. Lett. 98~(16)
  (2007) 160405.

\bibitem{Kantian:2009qv}
A.~Kantian, M.~Dalmonte, S.~Diehl, W.~Hofstetter, P.~Zoller, A.~J. Daley, Phys.
  Rev. Lett. 103~(24) (2009) 240401.

\bibitem{Seiberg:1984eu}
N.~Seiberg, Phys. Rev. Lett. 53~(7) (1984) 637--640.

\bibitem{Paredes08}
B.~Paredes, I.~Bloch, Phys. Rev. A 77 (2008) 023603.

\end{thebibliography}




\end{document}